\documentclass[a4paper,11pt]{article}
\usepackage{jinstpub} 
\usepackage{lineno}

\usepackage[freestanding]{hepunits}
\usepackage{siunitx}

\usepackage{ulem}

\usepackage{natbib}
\proceeding{Topical Workshop on Electronics for Particle Physics - TWEPP2025\\
Rethymno, Crete, Greece\\
Oct 6-10, 2025 }

\title{\boldmath The read-out electronics for the FLASH experiment}

\collaboration[c]{on behalf of FLASH collaboration}

\author[a,b,1]{L. Calligaris\note{Corresponding author.}}
\author[a]{C. Puglia}
\author[a,b]{G. Lamanna}
\affiliation[a]{Istituto Nazionale di Fisica Nucleare - Sezione di Pisa,\\
Largo Bruno Pontecorvo 3, Pisa, Italy}
\affiliation[b]{Dipartimento di Fisica, Università di Pisa,\\
Largo Bruno Pontecorvo 3, Pisa, Italy}

\emailAdd{luigi.calligaris@cern.ch}

\abstract{We introduce the FLASH haloscope experiment and present its electronic read-out system, currently under development. FLASH searches for Dark Matter (DM) particles and High-Frequency Gravitational Waves (HFGWs) using two cryogenic resonant cavities to scan the radio frequency spectrum between 117 and 360 MHz, looking for signals as weak as 10-22 W. The signal read-out uses Microstrip Superconducting Quantum Interference Amplifiers (MSAs) as low-noise amplifiers and Software-Defined Radio (SDR) techniques to acquire, preprocess and reduce the physics signal to a format suitable for permanent storage and offline analysis.}


\arxivnumber{2603.23846} 

\notoc

\begin{document}
\maketitle
\flushbottom
\newpage

\setcounter{page}{1}

\section{Introduction}
\label{sec:introduction}
\vspace{-3mm}
One of the most pressing problems of current research in the physics of the universe is the nature of dark matter (DM) \cite{Rubin:1970zza, Trimble:1987ee, Efstathiou:1992sy, Planck:2018nkj}, cold (non-relativistic) matter which gives strong hints on its existence in astrophysical and cosmological observations but has so far eluded observation. A model put forward to describe the composition of DM is the axion\cite{Weinberg:1977ma, Wilczek:1977pj, Peccei:1977ur, Peccei:1977hh}, a light pseudoscalar particle with feeble coupling to the photon. If cold DM has a significant component made of axions, planet Earth is traversing a steady stream of axions trapped in the gravitational well of the galaxy. In the current experimental landscape, the axion is being searched in the $m_{a} \simeq 10^{-7}~\text{to}~1~\eV/\text{c}^{2}$ mass range, with a coupling to the photon in the range $|g_{\gamma a}| = 10^{-17}~\text{to}~10^{-10}~\GeV^{-1}$.

The FINUDA magnet for Light Axion SearcH (FLASH) experiment will search for the QCD axion in the mass range $0.49~\text{to}~1.49~\mu\eV$, looking for conversion of axions into photons in the strong field of a superconducting magnet, exciting the electromagnetic modes of a resonant cavity. This type of detector is known as \textit{haloscope}. The axion mass range of interest corresponds to converted photons in the radio spectrum ($117~\text{to}~360$ MHz). While the detector is designed for an axion search, any feebly-interacting particle coupling to photons, such as axion-like particles\cite{Svrcek:2006yi, Khoury:2003aq, Arvanitaki:2009fg}, scalars and chameleons \cite{Khoury:2003aq}, can be probed by the experiment in the ranges of interest. A haloscope like FLASH, optimized for the aforementioned frequency range, is also sensitive to High-Frequency Gravitational Waves (HFGW) \cite{Aggarwal:2020olq}, which are predicted to be emitted by astrophysical processes such as the merger of light primordial black holes ($m_{\text{BH}} = 10^{-9}~to~10^{-1}~M_\odot$). The HFGWs couple to multiple resonant modes of the cavity, making it possible to monitor them concurrently and search for signals in coincidence. In FLASH, this coincidence concept is further extended by sharing the HFGW candidate data with other participants to the GravNet\cite{Schneemann:2024qli} project, which builds a global network of such detectors.

\section{Design of the experiment}
\label{sec:experiment_design}
\begin{figure}
    \centering
    \includegraphics[width=0.80\textwidth]{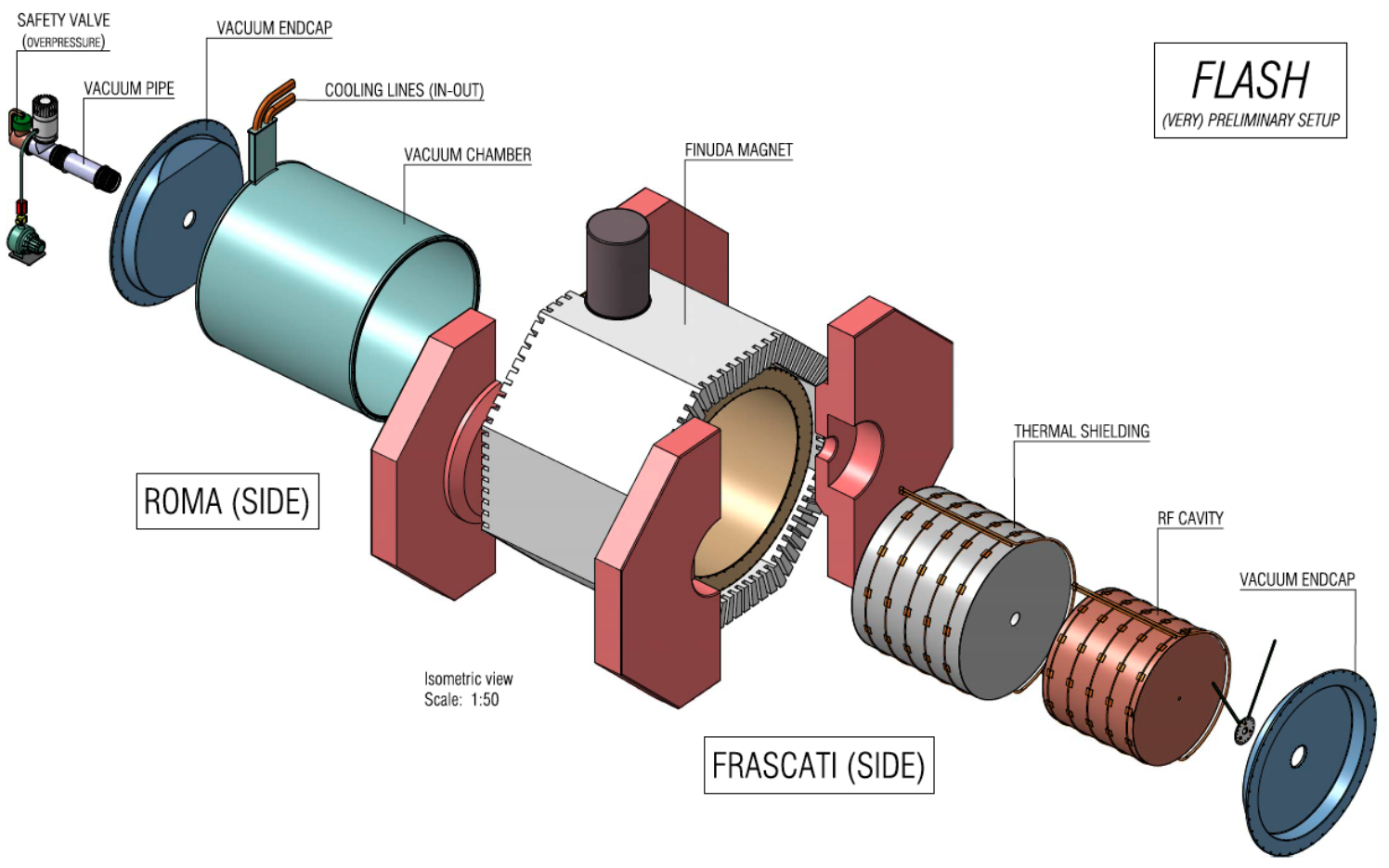}\\
    \vspace{5mm}
    \includegraphics[width=0.99\textwidth]{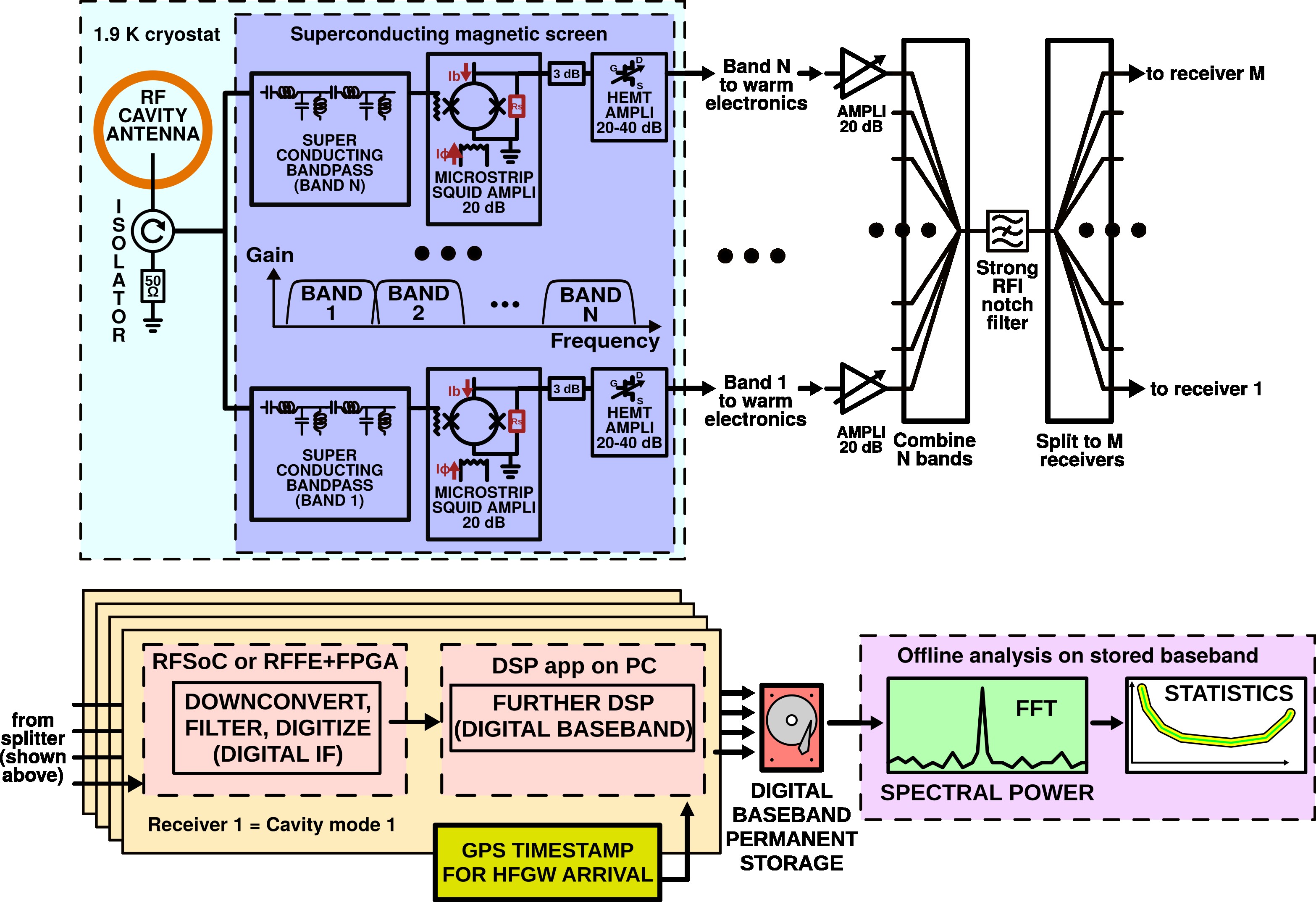}
    \caption{(top) Exploded view of the FLASH experiment, showing its main structural components. (bottom) Schematic of the readout chain for FLASH.}
    \label{fig:flash_layout}
\end{figure}

\vspace{-3mm}
The FLASH haloscope (shown in the top part of Fig.\ref{fig:flash_layout}) uses the former superconducting solenoid of the FINUDA \cite{FINUDA:1993xwu} collider experiment at INFN Laboratory Nazionali di Frascati, Italy, to generate a $1.1 ~\tesla$  magnetic field in a cylindrical volume hosting the resonant cavity. Two cylindrical cavities of equal length will be operated over the life of the experiment, a $1200x1050~\mm$ large cavity covering the $117 - 206~\MHz$ range and a smaller $1200x590~\mm$ one covering the $206 - 360~\MHz$ range for the TM010 mode used in the axion search. Higher order modes extending up to a frequency of around 900 MHz will be used in the HFGW search and are under investigation. The cavity is machined from oxygen-free high thermal conductivity copper (OFHC) and kept at a temperature of $1.9~\kelvin$, both to reduce thermal noise and improve its conductivity, such that the cavity quality factor reaches values of up to $Q = 5.78 \cdot 10^{5}$ for the TM010 mode. A set of cylindrical copper rods oriented along the axis of the cylindrical cavity and positioned inside it can be rotated outward or inwards, providing coarse tuning for the resonant frequency of the cavity, while a small (1 $\cm$ radius) alumina or sapphire rod can be inserted in the cavity through a port to provide fine tuning of the resonant frequency. At the lowest value for $|g_{\gamma a}| = 10^{-17}~\GeV^{-1}$ in the search parameters space, the power of photons converted from DM axions exciting the cavity is of the order of $10^{-22}~\watt$, which corresponds to $-190~\decibel \text{m}$. The read-out electronics (shown in the bottom part of Fig. \ref{fig:flash_layout}) are designed to be able to detect such a power level above the noise floor by integrating the signal power corresponding to each frequency step over a time of 5 to 10 minutes.

\section{Overview of the read-out chain}
\label{sec:overview_of_the_readout_chain}
\vspace{-3mm}
A small coaxial non-resonant antenna couples to the cavity modes by being inserted into the cavity volume. This antenna is electrically short (few mm to 20 mm) compared to the wavelength and its length is a compromise between reading out a measureable signal and not degrading excessively the quality factor of the cavity, which should stay in the range $3 \cdot 10^{5}$ to $7 \cdot 10^{5}$. The signal proceeds through a 50 $\ohm$ coaxial line to an isolator, preventing RF energy to be fed back into the cavity, and then to a set of band-pass filters, which split the whole range of frequencies tunable by the cavity into bands matching the bandwidth of the individual first stage amplifiers and reject noise outside of the bandwidth of interest for the measurement. This splitting is chosen because in the specifications provided by the manufacturer, the kind of first-stage amplifier we intend to use (described below) exhibits a 3dB-bandwidth of around 100 MHz centered around a fixed frequency, which is not sufficient to cover the entire range of frequency corresponding to the modes of interest for the search of axions and HFGWs. Thus, there will be as many first-stage amplifiers and filters as needed to cover the frequency range of interest, which goes from 117 MHz to several hundred MHz (order of 600-900 MHz, a precise upper frequency limit depends on which modes are chosen to perform the HFGW search, which is still under study). 
The first stage amplifiers need to be characterized by low noise (order of tens of mK) to guarantee an instantaneous signal-to-noise ratio in the amplification chain such that signal integration time for the estimation of power spectrum can be kept on the order of 5-10 minutes per frequency step. To accommodate this low-noise requirement, Microstrip SQUID Amplifiers (MSA)\cite{10.1063/1.121490} - described below - have been chosen for the application. These devices can achieve between 38 mK and 118 mK noise temperature across the FLASH frequency range, which corresponds to around 7 times the quantum limit. Following each first-stage amplifier is a second stage amplifier based on a High Electron Mobility Transistor (HEMT). Both first- and second-stage amplifiers are cooled at the common temperature of the cavity ($1.9~\kelvin$) to reduce their thermal noise. The MSA and the HEMT amplifiers are expected to exhibit a gain of around $20\dB$ and $40\dB$ (tunable), respectively.

\section{Superconducting band-pass filters}
\label{sec:superconducting_band_pass_filters}
\vspace{-3mm}
The band-pass filters need to be operated at $1.9~\kelvin$, both to keep thermal noise under control and because the devices at their edges are operating at the same temperature. The realization of the filters with a traditional PCB populated with discrete components is an option, but would entail subjecting the components to an important thermal stress as the cryostat is cycled between ambient and operating temperature during maintenance and technical stops. The use of normal-conducting components would also lead to some signal attenuation due to ohmic losses, which is to be avoided before the signal is amplified by the MSA. In view of the motivations above, we looked into a type of filter which could exploit the available low temperatures in the cryostat, designing a set of band-pass filters built from superconducting material. This material offers two advantages: it does not attenuate the signal through ohmic losses and allows to realize the inductive components using kinetic inductance \cite{10.1007/s10909-007-9685-2}, which makes the device more compact and less magnetically coupled to external noise. The filters are implemented as a chain of lumped elements manufactured using thin superconducting films on a silicon substrate
\cite{US20110152104A1,longobardi2013microstrip,6373700}, using the same lithographic technology used in the manufacture of cryogenic sensors such as Transition-Edge Sensors (TES) or Kinetic Inductance Detectors (KIDs). The filters are designed to be manufactured using niobium or niobium alloys, since these materials are superconducting at the operating temperature of the experiment, their technology is industrially well-established in terms of supplies and tooling and, lastly, the facilities available at our institutes support their use in the manufacturing of prototype devices.

\section{Microstrip SQUID Amplifier}
\label{sec:microstrip_squid_amplifier}
\vspace{-3mm}
The MSA is a Superconducting Quantum Interference Device (SQUID), in which the magnetic flux associated to the input signal is coupled to the superconducting loop of the amplifier through a resonant microstrip antenna. The device can be manufactured to operate at frequencies of several hundreds of MHz, making it suitable for our application. We expect to characterize in 2026 two specimens of recently-manufactured MSAs, of a kind which would allow to cover the spectrum of interest with 5-8 of these amplifiers. 

\section{Warm amplification, signal digitization and processing}
\label{sec:signal_digitization_and_processing}
\vspace{-3mm}
Following amplification in the two cryogenic stages, the signal passes through a low-thermal-conductivity coaxial cable and is brought out of the cryostat. There, a commercial amplifier with a gain of $20\dB$ brings the signal to a level which can be acquired by a commercial Software-Defined Radio (SDR) device. The signals from different bands will be combined and filtered from strong radio frequency interference (RFI), such as the FM and DAB commercial radio and mobile phone bands, to be again split to feed the RF receivers, each of them tracking a cavity mode. Two classes of SDR devices have been considered for the application: direct RF conversion devices and zero-IF RF front-ends.

\begin{figure}
    \centering
    \includegraphics[height=3.5cm]{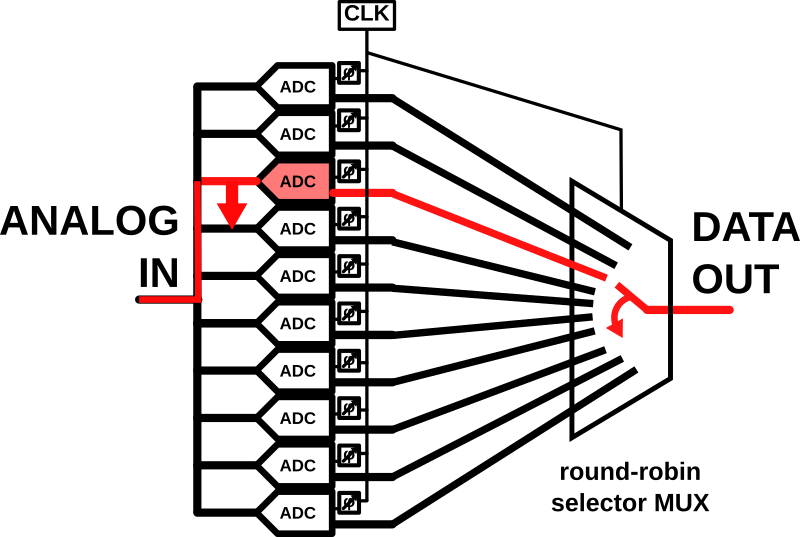}
    \hfill
    \includegraphics[height=2.0cm]{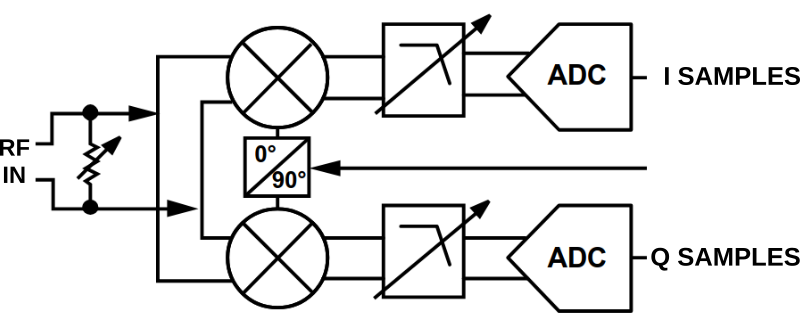}
    \hfill
    \includegraphics[height=2.3cm]{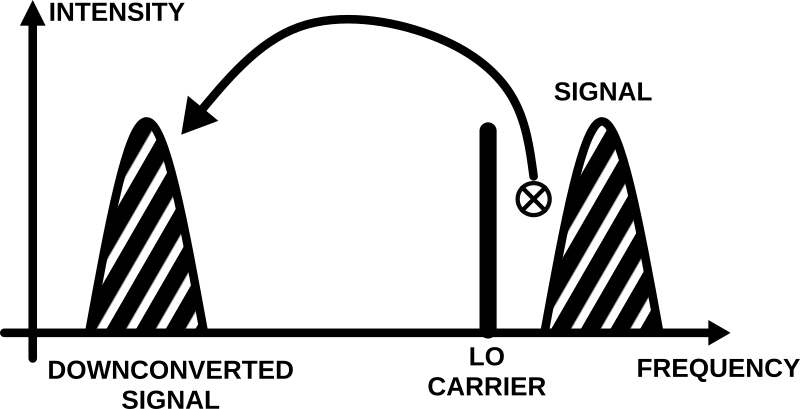}
    \caption{(left) Schematic of a composite ADC. (center) Schematic of a zero-IF receiver. (right) Representation of downconversion of a signal in frequency space.}
    \label{fig:rf_frontends}
\end{figure}

In a direct RF conversion device the waveform is acquired directly without down-conversion, usually at a sampling rate suitable to keep the signal into the 1st Nyquist zone, that is, a frequency at least twice the maximum frequency among the components of the signal of interest. The device we considered for this category is the AMD Zynq Ultrascale+ RFSoC, an FPGA array coupled with a hard processing system and high-speed ADCs and DACs, capable of acquiring signals at a maximum rate of 5 billion samples per second (5 GSPS). To achieve such a very high sampling speed the ADCs are realized as a composite ADC (shown in Fig.\ref{fig:rf_frontends}), that is, an array of identical ADCs running at a lower sampling speed (of the order of 500 MSPS) which are triggered in sequence by a tunable phasing network connected to the main sampling clock. The use of this kind of ADC gives a significant benefit in terms of simplicity in the processing of the signal, since the very wide acquisition bandwidth means that the whole frequency spectrum associated to the cavity can be acquired and then digitally processed with FPGA algorithms in the same device. The drawbacks lie in the fact that the ADC is exposed to signals in the entire acquisition range, leading to the sampling of noise outside of the frequency band of interest and therefore a reduced dynamic range for the signal. Additionally, small phasing and gain errors in the individual ADCs inside the composite ADC introduce spurs in the acquired RF spectrum, polluting with artifacts the signal region.

In a zero-IF device, the RF signal is mixed with a sinusoidal carrier, leading to the appearance of frequency sum and frequency difference signals between the input signal and the carrier. The frequency of the carrier is chosen such that the difference signal lies at a low frequency, such that it can be selected using a low-pass filter, this whole process being referred to as \textit{downconversion}.  Following the filter an ADC samples the downconverted signal, making it available for digital processing by an FPGA. The devices of this kind being considered for the application are two RF front-ends from Analog Devices, the AD9361 and the ADRV9002. Both the devices integrate the downconversion and ADC hardware in a single package, together with programmable analog filters. The AD9361 features two ADCs, allowing to acquire two channels with a single device, but a single local oscillator (LO), which limits the flexibility in tuning independently the two receive channels. In this respect the ADRV9002 is favoured in the choice between the two, since it features two independent LOs, which gives the possibility to independently tune the two receive channels, and thus individually track two cavity modes as the tuning bars are rotated. The use of a zero-IF device allows to select a narrow window in frequency space around the downconversion carrier and, therefore, limits the amount of out-of-band noise and signals entering the ADC. Being equipped with a high quality downconversion carrier oscillator, these RF front-ends feature a relatively clean and artifact-free spectrum in the acquisition window, with the possible exception of a spike at zero frequency due to a common DC offset in the signal entering the ADC, which can be easily managed with simple filtering.

Five commercial-off-the-shelf solutions based on the devices described above are being procured for characterization: the Zynq Ultrascale+ RFSoC will be evaluated using a \textit{RealDigital RFSoC 4x2} board; the AD9361 will be evaluated using either an \textit{Ettus Research USRP B210} SDR or an \textit{Analog Devices AD-FMCOMMS3-EBZ} FMC daughterboard coupled with a Zynq Ultrascale+ development board manufactured by Trenz Electronics (model TE0818-02-9BE81-AS); the ADRV9002 will be evaluated using either an \textit{Analog Devices JupiterSDR} or an \textit{Analog Devices EVAL-ADRV9002} FMC daughterboard coupled with the aforementioned Trenz Electronics Zynq development board. In our testing, we will evaluate the simplicity of use of ready-integrated solutions as the B210 and the JupiterSDR against the flexibility of programming the FPGA in the  RFSoC 4x2, AD-FMCOMMS3-EBZ and EVAL-ADRV9002 options with custom digital signal processing algorithms. We will also compare the performance of the direct RF sampling and zero-IF schemes in detecting weak signals, which will be emulated by generating them with instruments such as function generators and SDRs and passed through a chain of attenuators held at cryogenic temperatures.

\vspace{-4mm}
\section{Conclusions}
\label{sec:conclusions}
\vspace{-3mm}
We introduced the FLASH experiment and described the  electronics of its read-out system. The experiment requires tracking a number of resonant modes in a high-quality copper cavity, looking for extremely tiny signals (down to $10^{-22}$ W) using a low-noise acquisition chain. The first- and second-stage analog amplifiers use very low-noise cryogenic MSA and HEMT technology, respectively, while the later amplification and digitization stages use room temperature, commercial-off-the-shelf software-defined-radio devices which will be characterized soon.

\acknowledgments

We want to thank Claudio Gatti for his steadfast support as principal investigator of the FLASH experiment. FLASH is funded by the European Union (GravNet, No. ERC-2024-SyG 101167211, DOI: 10.3030/101167211).

\bibliography{biblio.bib}

\end{document}